\documentstyle[12pt,twoside,fleqn,espcrc1,epsfig]{article}

\newcommand{\AmS}{{\protect\the\textfont2
  A\kern-.1667em\lower.5ex\hbox{M}\kern-.125emS}}

\title{Phenomenological aspects of kaon photoproduction
       on the nucleon}

\author{T. Mart\address{Jurusan Fisika, FMIPA, Universitas Indonesia,
        Depok 16424, Indonesia}\thanks{Supported in part by the 
        University Research for Graduate Education
        (URGE) grant.}, S. Sumowidagdo$^{\rm a*}$, 
        C. Bennhold\address{Center for Nuclear Studies, Department 
        of Physics, The George Washington University, \\ 
        ~\,Washington, D.C. 20052, USA}\thanks{Supported in part by
        US DOE with grant no. DE-FG02-95ER-40907},
        and H. Haberzettl$^{\rm b\dagger}$}

\begin{document}
\maketitle

\begin{abstract}
Using an isobar model which can reproduce the existing experimental
data of kaon photoproduction on the nucleon we investigate some 
related phenomenological aspects, i.e. the hadronic form factors, 
contributions of kaon-hyperon final states to the anomalous 
magnetic moment of the nucleon, and missing nucleon resonances. 
By reggeizing the appropriate propagators we extend the model to 
the higher photon energy regime.
\end{abstract}

\section{INTRODUCTION}

A wealth of new high-statistics data on elementary kaon photo- and
electroproduction has recently become available in three isospin 
channels. Along with some new progress in the theoretical side this
has made the field of kaon electromagnetic production to be of
considerable interest. New models, span from chiral perturbation
theory to the relatively simple isobar approach, have been proposed
in the recent years as the {\footnotesize SAPHIR} collaboration made 
their precise data publicly available. Because the error-bars are 
sufficiently small, an interesting structure can be resolved in 
$K^+\Lambda$ total cross section. This leads to a critical question,
as to whether the structure comes from less known resonances or
other reaction channels start to open at the corresponding energy.

In this paper we discuss some phenomenological aspects, which can
be investigated by means of the isobar model. The model has been
constructed by including three states that have been found to have 
significant decay widths into $K\Lambda$ and $K\Sigma$ channels, 
the $S_{11}$(1650), $P_{11}$(1710), and $P_{13}(1720)$ resonances, 
to fit all elementary data by adjusting some free
parameters, which are known as coupling constants.

\section{THE MODEL AND SOME PHENOMENOLOGICAL ASPECTS}
\subsection{Hadronic Form Factors}
Previous analyses of kaon photoproduction have never included a 
form factor at the hadronic vertex. However, since most of the 
present isobaric models diverge at higher energies, the need for 
such hadronic form factors has been known for a long time. 
Furthermore, it has been demonstrated that models which give 
a good description of the $(\gamma,K^+)$ data can give unrealistically 
large predictions for the $(\gamma,K^0)$ channels \cite{terry95}.
It is well known that incorporating a hadronic form factor helps alleviate 
this divergence and, simultaneously, leads to a problem with gauge 
invariance, since not every diagram in the Born terms retains gauge 
invariance by itself.\footnote{Since the resonance terms are individually 
gauge invariant, the discussion will be limited to
the Born terms.} The question of gauge invariance is 
actually one of the central issues in dynamical descriptions of 
how photons interact with hadronic systems. While there is usually no 
problem at the tree-level with bare, point-like particles, the problem 
becomes very complicated once the electromagnetic interaction is 
consistently incorporated within the full complexity of a 
strongly-interacting hadronic system.

In the previous work \cite{bennhold97} we have studied the 
influence of hadronic form factors on kaon production, by multiplying the 
whole amplitude with an overall, monopole, form factor $F(t)$, i.e.
\begin{eqnarray}
M_{\rm fi} &=& [\, M_{\rm B}(s,t,u) ~+~ M_{\rm R}(s,t,u)\, ] \cdot F(t) ~,
\end{eqnarray}
where the subscripts B and R refer to the Born and resonance terms, 
to simulate the average effect of the fact that nucleons are not point-like.
In spite of the success to suppress the divergence and to avoid the problem 
of gauge invariance, this {\it ad hoc} fashion does not have any microscopic 
foundation.

In order to restore gauge invariance properly, one needs to construct 
additional current contributions beyond the usual Feynman diagrams to 
cancel the gauge-violating terms. One of the most widely used methods 
is due to Ohta \cite{ohta89}. For kaon photoproduction off 
the nucleon, Ohta's prescription amounts to dropping all 
strong-interaction form factors for all gauge-violating electric current 
contributions in Born terms. Symbolically, this may be written as
\begin{eqnarray}
M_{\rm B}(s,t,u) &=& M^{\rm mag.}_{\rm B}\,[\, s,t,u,F(s),F(t),F(u)\, ] ~+~ 
M^{\rm elec.}_{\rm B}(s,t,u) ~.
\end{eqnarray}
The recipe, however, does not completely solve the problem of divergence,
since the electric terms do not have suppression and, therefore, 
could violently increase as a function of the coupling constants. 
As shown in Ref.\,\cite{hbmf98}, even at the coupling constants 
values accepted by the SU(3) symmetry, Ohta's recipe already yields
very large $\chi^2$.

On the other hand, Haberzettl \cite{hh97} has put forward a comprehensive 
treatment of gauge invariance in meson photoproduction. This includes a 
prescription for restoring gauge invariance in situations when one cannot
handle the full complexity of the problem and therefore must resort 
to some approximations. In our language, this method can be translated as
\begin{eqnarray}
M_{\rm B}(s,t,u) &=& M^{\rm mag.}_{\rm B}\,[\, s,t,u,F(s),F(t),F(u)\, ] ~+~ 
M^{\rm elec.}_{\rm B}(s,t,u) \cdot {\hat F}(s,t,u) ~,
\end{eqnarray}
with ${\hat F}(s,t,u) = a_1F(s)+a_2F(t)+a_3F(u)$ and  $a_1+a_2+a_3=1$.
Clearly, Haberzettl's method removes the Ohta's problem by an additional
form factor in the electric terms.

By fitting to the kaon photoproduction data we found that the method
proposed by Haberzettl to be superior rather than the Ohta's, since the  
former can provide a reasonable description of the data using values 
for the leading couplings constants close to the SU(3) prediction.
Such couplings cannot be accommodated in Ohta's method due to the 
absence of a hadronic form factor in the electric current contribution.

\subsection{The anomalous magnetic moment of the nucleon}
One of the important ground state properties of the nucleon is
the anomalous magnetic moment, which exists as a direct consequence
of its internal structure. More than 30 years ago Gerasimov, and
independently Drell and Hearn, proposed that this ground state
property is related to the nucleon's resonance spectra 
by a sum rule which was then called the Gerasimov-Drell-Hearn 
({\footnotesize GDH})
sum rule \cite{gdh}. In the limit of photon point, the sum rule may 
be written as
\begin{eqnarray}
  \label{eq:gdh1}
  -\frac{\kappa_N^2}{4} &=& \frac{m_N^2}{8\pi^2\alpha}\,\int_0^\infty \,
  \frac{d\nu}{\nu} ~ [ \sigma_{1/2}(\nu)-\sigma_{3/2}(\nu)] ~,
\end{eqnarray}
where $\sigma_{3/2}$ and $\sigma_{1/2}$ denote the cross sections for 
possible combinations of the nucleon and photon spins.
Experiment with polarized beam and target has been performed at 
{\footnotesize MAMI} with photon energy up to 850 MeV 
and data are being analyzed \cite{mainz}. Using higher photon 
energies, experiments have been planned  at 
{\footnotesize ELSA} and {\footnotesize JLab}.

\begin{table}[t!]
\newlength{\digitwidth} \settowidth{\digitwidth}{\rm 0}
\caption{Numerical values for the contribution of kaon-hyperon final
      states to the square of  
      anomalous magnetic moments of proton and neutron. 
      Column (1) is obtained from Eq.\,(\ref{eq:gdh2}), while column (2) 
      is evaluated by using Eq.\,(\ref{eq:gdh3}).
      Experimentally, $\kappa_p^2=3.214$ and $\kappa_n^2=3.660$.}
\renewcommand{\arraystretch}{1.2}
\label{tab:gdh}
\begin{tabular*}{\textwidth}{@{}l@{\extracolsep{\fill}
                                   }rr@{}l@{\extracolsep{\fill}}rr}
\hline
                 & \multicolumn{2}{c}{$\kappa_p^2(K)$} & 
                 & \multicolumn{2}{c}{$\kappa_n^2(K)$} \\
\cline{2-3} \cline{5-6}
Channel          & \multicolumn{1}{c}{(1)}
                 & \multicolumn{1}{c}{(2)} &
Channel          & \multicolumn{1}{c}{(1)}
                 & \multicolumn{1}{c}{(2)}         \\
\hline
$\gamma\; p \rightarrow K^+ \Lambda$&$-0.026$&$0.044$&$\gamma\; 
n \rightarrow K^0 \Lambda$&$0.075$&$0.110$\\
$\gamma\; p \rightarrow K^+ \Sigma^0$&$-0.024$&$0.030$&$\gamma\; 
n \rightarrow K^+ \Sigma^-$&$-0.025$&$0.050$\\
$\gamma\; p \rightarrow K^0 \Sigma^+$&$-0.013$&$0.031$&$\gamma\; 
n \rightarrow K^0 \Sigma^0$&$-0.019$&$0.031$\\ 
Total&$-0.063$&$0.105$&Total&$0.031$&$0.191$\\
\hline
\end{tabular*}
\end{table}

For practical purpose, instead of Eq.\,(\ref{eq:gdh1})  we use
\begin{eqnarray}
  \label{eq:gdh2}
  {\kappa_N^2} &=& \frac{m_N^2}{\pi^2\alpha}
  \int_{\nu_{\rm thr}}^{\nu_{\rm max}} \frac{d\nu}{\nu}\; \sigma_{TT'} ~,
\end{eqnarray}
where $\sigma_{TT'}$ denotes the cross section with polarized real
photon and target. In terms of polarization observables this cross 
section corresponds to the double polarization $E$ \cite{workman92}.
Since there are no data available for $\sigma_{TT'}$, previous work 
\cite{drechsel} approximated Eq.\,(\ref{eq:gdh2}) with
\begin{eqnarray}
  \label{eq:gdh3}
  {\kappa_N^2} &\leq&\frac{m_N^2}{\pi^2\alpha}
  \int_{\nu_{\rm thr}}^{\nu_{\rm max}} \frac{d\nu}{\nu}\; \sigma_{T} ~,
\end{eqnarray}
to estimate the upper bound of contributions, where $\sigma_T$ represents
the total cross section.

To calculate Eqs.\,(\ref{eq:gdh2}) and (\ref{eq:gdh3}) we use our
elementary operator with $\nu_{\rm max}=2.2$ GeV. 
The result is shown in Table~\ref{tab:gdh}.  
Our calculation yields  values of $\kappa_p^2(K)= -0.063$ and 
$\kappa_n^2(K)= 0.031$, or $|\kappa_p(K)|/\kappa_p \leq 0.14$ and 
$\kappa_n(K)/\kappa_n \leq 0.094$.  
This shows that the kaon-hyperon final states contributions to the 
proton's and neutron's magnetic moment are very small. An
interesting feature is that our calculation yields a negative contribution
for the  $\kappa^2_p(K)$ and a positive contribution for the $\kappa^2_n(K)$,
which is obviously consistent with the result of Karliner's work
\cite{karliner73}.

\subsection{Investigation of missing resonances}

\begin{figure}[!t]
\begin{minipage}[t]{105mm}
\epsfig{file=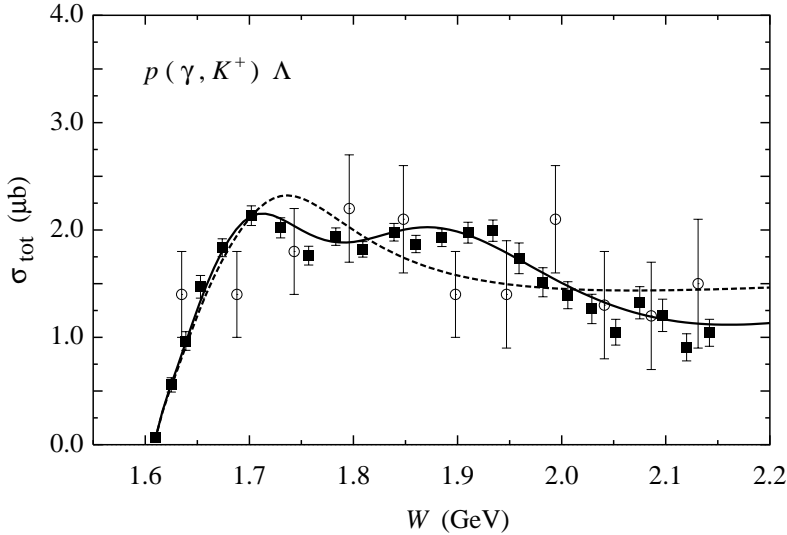, width=100mm}
\end{minipage}
\hspace{\fill}
\begin{minipage}[!t]{50mm}
\vspace{-9.2cm}
\caption{Total cross section for $K^+\Lambda$ photoproduction
         on the proton. The dashed line shows the model without the 
         $D_{13}(1960)$ resonance, while the solid line is obtained 
         by including the $D_{13}(1960)$ state. 
         The new {\footnotesize SAPHIR} data 
         \cite{saphir98} are denoted by the solid squares, old data  
         are shown by the open circles.}
\label{fig:missing}
\end{minipage}
\vspace{-0.6cm}
\end{figure}

A brief inspection to the particle data book reveals that less than 40\% 
of the predicted nucleon resonances are observed in 
$\pi N\to \pi N$ scattering experiments. Quark model studies have 
suggested that those "missing" resonances may couple strongly to 
other channels, such as the $K \Lambda$ and $K \Sigma$ channels 
\cite{capstick98}. Interestingly, 
the new {\footnotesize SAPHIR} total cross section data \cite{saphir98} for 
the $p(\gamma, K^+)\Lambda$ channel, shown in Fig.\,\ref{fig:missing},
indicate for the first time a structure around $W = 1900$ MeV. 
Using the current isobar model we investigate this 
structure. As shown in Fig.~\ref{fig:missing}, our previous model 
cannot reproduce the total cross section. Although a structure in 
total cross section data does not immediately imply a new resonance, 
the energy region around 1900 MeV represents a challenge not only 
because of possible broad, overlapping resonances, but also because 
there are additional production thresholds nearby, such as 
photoproduction of $\eta' $, $K^* \Lambda$, and 
$K \Lambda^*$ final states, which can all lead to structure in the 
$K^+ \Lambda$ cross section through final-state interaction.
Here, we limit ourselves only to the possibility that this 
structure is in fact due to one of the missing or poorly known resonances.

The constituent quark model of Capstick and Roberts \cite{capstick98} 
predicts many new states around 1900 MeV. However, only a few of them have 
been calculated to have a significant $K \Lambda$ decay width 
\cite{capstick98}. These are the $S_{11}$(1945), $P_{11}$(1975), 
$P_{13}$(1950), and $D_{13}$(1960) states. We have performed fits for each 
of these possible states, allowing the fit to determine the mass, width 
and coupling constants  of the resonance. We found that all four states 
can reproduce the structure at $W$ around 1900 MeV, while reducing the 
$\chi^2$, but only for the  $D_{13}$(1960) state we found a remarkable 
agreement, up to the sign, between the quark model prediction and our 
extracted result \cite{terry2000}.
The result is shown in Fig.\,\ref{fig:missing}, where 
without this resonance the model shows only one peak
near threshold, while inclusion of the new resonance leads
to a second peak at $W$ slightly below 1900 MeV,
in accordance with the new {\footnotesize SAPHIR} data.
The difference between the two calculations is much smaller for 
the differential cross sections.
The largest effects are found in the photon asymmetry.
Therefore, we would suggest that measuring this observable is 
well suited to shed more light on the contribution of this state 
in kaon photoproduction.

\section{EXTENSION TO HIGHER ENERGIES}
\begin{figure}[!t]
\begin{minipage}[t]{90mm}
\epsfig{file=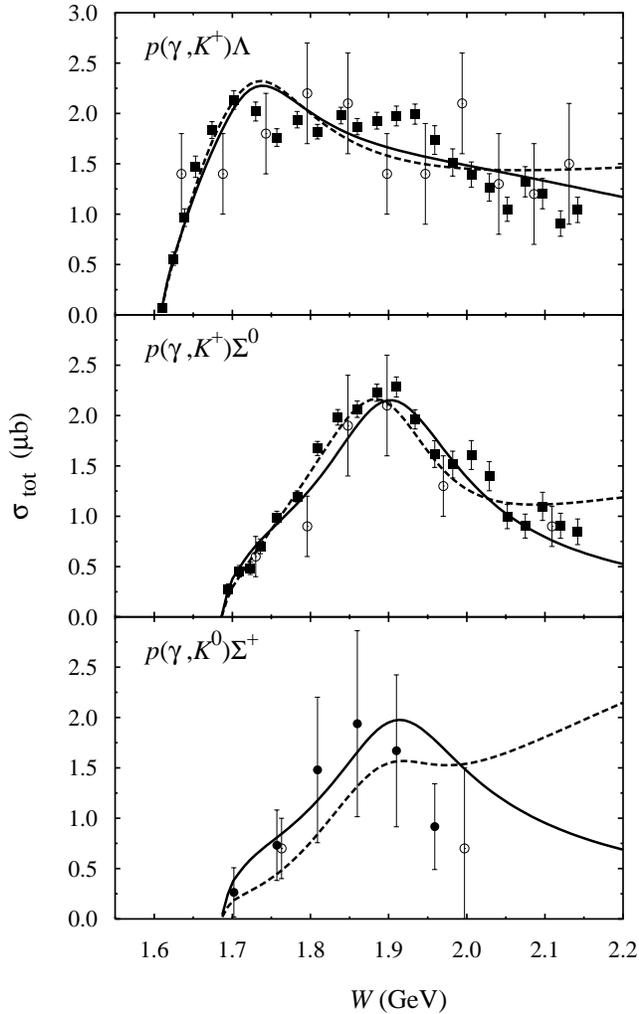, width=90mm}
\end{minipage}
\hspace{\fill}
\begin{minipage}[!t]{65mm}
\vspace{-10cm}
\caption{Total cross section for kaon photoproduction on the proton. 
         The dashed line shows the isobar model with hadronic form
         factors, but without reggeization. The solid line is obtained 
         by reggeizing the $K$ and $K^*$ propagators in the model. For $K^+$ 
         photoproduction, notation for the data is as in 
         Fig.\,\ref{fig:missing}, for $K^0$ production old (open circles)
         and preliminary data (solid circles) are shown \cite{frank99}.}
\label{fig:regge}
\end{minipage}
\vspace{-0.6cm}
\end{figure}

Extending the model to the higher energy regime requires a non-trivial task,
since the Born terms increase rapidly as a function of energy. As shown in
Fig.\,\ref{fig:regge}, even the hadronic form factors are unable to 
suppress the cross sections for the energy region above 2 GeV as
demanded by the data. Especially in the case of $K^0\Sigma^+$ production,
where the predicted cross section starts to monotonically increase at 
this point. However, in order to explore the higher-lying nucleon 
resonances or to account for higher energies contributions to the 
{\footnotesize GDH} integral, 
an isobar model which also properly work at higher photon energies
would be demanded.

In Ref.\,\cite{frank99} it has been shown that the contributions 
from the $t$-channel resonances are responsible for the divergence
of the cross section, thus 
indicating that the Regge propagator should be used instead of the 
usual Feynman propagator. While a proper reggeization of the model 
is considerably complicated and the study is still underway, we 
investigate here only the qualitative effects of using Regge 
propagators in the model. 

Following Ref.\,\cite{guidal97},  
we multiply the Feynman propagators $1/(t-m_{K^*}^2)$  of the 
$K^*(892)$ and $K_1(1270)$ resonances in the operator with a factor of 
$P_{\rm Regge}\cdot (t-m_{K^*}^2)$, where $P_{\rm Regge}$ indicates the
Regge propagator given in Ref.\,\cite{guidal97}. For the $K^*$ intermediate
state it has the form
\begin{eqnarray}
  \label{eq:regge}
  P_{\rm Regge} &=& \frac{s^{\alpha_{K^*}(t)-1}}{\sin [\pi\alpha_{K^*}(t)]}
                    ~ e^{-i\pi\alpha_{K^*}(t)} ~ 
                    \frac{\pi\alpha_{K^*}'}{\Gamma [\pi\alpha_{K^*}(t)]} ~,
\end{eqnarray}
where $\alpha (t) = \alpha_0 + \alpha '\, t$ denotes the corresponding
trajectory. Equation (\ref{eq:regge}) clearly reduces to the Feynman
propagator in the limit of $t \to m_{K^*}^2$, thus approximating the low
energy behavior of the amplitude.

The model is then refitted to kaon photoproduction data and 
the result is shown in Fig.\,\ref{fig:regge}, where we compare the
isobar model with and without reggeization. Obviously, Regge propagators
strongly suppress the cross section at high energies and, therefore, 
yield a better explanation of data at this energy regime. 
For the $K^0\Sigma^+$ process, the use of Regge propagators 
seems to give more flexibility in reproducing the cross section data. 
This cannot be achieved without 
reggeization, since the high energy behavior of both $t$-channel 
resonances is less controllable by the hadronic form factors. However, since 
the data for the $K^0\Sigma^+$ channel shown in Fig.\,\ref{fig:regge}
are still preliminary \cite{saphir99}, we have to wait before any further 
conclusion can be drawn. In future we will include the high energy data 
in the fit and investigate the model in the transition between medium 
and high energy regions.

\end{document}